\documentclass[aps,prl,twocolumn,reprint,showpacs,superscriptaddress]{revtex4-1}
\usepackage{amsmath}
\usepackage{amssymb}
\usepackage{graphicx}
\usepackage{hyperref}
\usepackage{color,xcolor}
\begin{document}
\title{Inversion of ferrimagnetic magnetization by ferroelectric switching via a novel magnetoelectric coupling}
\author{Yakui Weng}
\author{Lingfang Lin}
\affiliation{Department of Physics \& Jiangsu Key Laboratory for Advanced Metallic Materials, Southeast University, Nanjing 211189, China}
\author{Elbio Dagotto}
\affiliation{Department of Physics and Astronomy, University of Tennessee, Knoxville, TN 37996, USA}
\affiliation{Materials Science and Technology Division, Oak Ridge National Laboratory, Oak Ridge, TN 37831, USA}
\author{Shuai Dong}
\email{Corresponding author: sdong@seu.edu.cn}
\affiliation{Department of Physics \& Jiangsu Key Laboratory for Advanced Metallic Materials, Southeast University, Nanjing 211189, China}

\date{\today}

\begin{abstract}
Although several multiferroic materials/heterostructures have been extensively studied, finding strong magnetoelectric couplings for the electric field control of the magnetization remains challenging. Here, a novel interfacial magnetoelectric coupling based on three components (ferroelectric dipole, magnetic moment, and antiferromagnetic order) is analytically formulated. As an extension of carrier-mediated magnetoelectricity, the new coupling is shown to induce an electric-magnetic hysteresis loop. Realizations employing BiFeO$_3$ bilayers grown along the [$111$] axis are proposed. Without involving magnetic phase transitions, the magnetization orientation can be switched by the carrier modulation driven by the field effect, as confirmed using first-principles calculations.
\end{abstract}
\pacs{77.55.Nv, 73.21.-b, 75.70.Cn}
\maketitle

\textit{Introduction}.- Magnetoelectric (ME) effects and multiferroic materials are very important both for basic science and for practical applications \cite{Cheong:Nm,Ramesh:Nm,Dong:Ap}. However, to realize multiferroics into concrete devices, there are several crucial physical issues still to be addressed. Not only the ferroic properties, e.g. the ferroic phase transition temperatures ($T_{\rm C}$'s), magnetization ($\textbf{M}$), and polarization ($\textbf{P}$), must be increased \cite{Cheong:Nm,Dong:Mplb}, but also the coupling strength between spin moments and charge dipoles should be intrinsically stronger. Although a few exceptional multiferroic materials, such as BiFeO$_3$ and its heterostructures, show promising properties \cite{Heron:Apr,Yu:Ptrsa,Lu:Apr,Huang:Mplb}, several improvements are still required to achieve direct and effective ME functions at room temperature, especially to obtain an electric-magnetic ($\textbf{E}$-$\textbf{M}$) hysteresis loop.

Phenomenologically, any magnetoelectric energy term can be expressed as a function of the ferroic moments, $\textbf{P}$ and $\textbf{M}$, satisfying the energy symmetry requirement that they transform as a scalar \cite{Dong:Ap}. For example, the most canonical one, $\textbf{P}^{2}\textbf{M}^{2}$, represents pure strain mediated magnetoelectricity that often occurs in composites consisting of simple piezoelectric and magnetostric components. One of the most important recent achievements in multiferroics is the discovery of several other ME mechanisms beyond this simple $\textbf{P}^{2}\textbf{M}^{2}$. For example, a complex interaction term $\textbf{P}\cdot[\textbf{M}(\nabla\cdot\textbf{M})-(\textbf{M}\cdot\nabla)\textbf{M}]$ was proposed \cite{Mostovoy:Prl06}, which is associated with the Dzyaloshinskii-Moriya interaction (or spin current) mediated ME coupling in spiral magnets \cite{Katsura:Prl,Sergienko:Prb}.

In heterostructures, there are many other possibilities. For example, the field-effect ME coupling can be carrier-mediated in heterostructures involving ferroelectrics (or dielectrics) and ferromagnets \cite{Rondinelli:Nn,Duan:Prl,Huang:Mplb,Vaz:Jpcm,Fechner:Prb}, and it can be expressed as $(\nabla\cdot\textbf{P})\textbf{M}^{2}$ (or $(\nabla\cdot\textbf{P})|\textbf{M}|$). Microscopically, the magnetic response to electric fields is achieved by accumulating or depleting carriers (electrons or holes) near the interface via the field effect \cite{Huang:Mplb,Vaz:Jpcm}. In this case, the sign of $\textbf{M}$ can not be switched but its amplitude ($|\textbf{M}|$) can be tuned because is proportional to the carrier density. For correlated electronic systems, magnetic phase transitions can be obtained upon carrier modulation, which may amplify this carrier-mediated ME response \cite{Vaz:Prl,Molegraaf:Am,Burton:Prb,Burton:Prl,Dong:Prb11,Chen:Prb12,Yin:Nm,Jiang:Nl,Dong:Prb13.2}.
Despite the considerable modulation of $|\textbf{M}|$, the sign of $\textbf{M}$ is still not switchable upon electric switching. Furthermore, magnetic phase transitions are not easy to control in real experiments because the system must be fine tuned to be located near phase boundaries. Realizing sensitive ME responses based on phase transitions of robust magnetic states remains a challenge.

\textit{New ME coupling: $(\nabla\cdot\textbf{P})(\textbf{M}\cdot\textbf{L})$.}- In this publication, a new mechanism for ME coupling will be proposed based on the carrier-mediated field effect. This novel coupling does not depend on magnetic phase transitions and it can lead to a $180^\circ$ switching of $\textbf{M}$. The key observation is to replace $\textbf{M}^{2}$ in the aforementioned formula by $\textbf{M}\cdot\textbf{L}$, where $\textbf{L}$ is the AFM order parameter. In the presence of robust AFM order (i.e. robust $\textbf{L}$), the direction of $\textbf{M}$ can be switched accompanying the switching of $\textbf{P}$.

How to realize this new ME coupling in real materials? In general, the field effect, in the form of $\nabla\cdot\textbf{P}$, is layer dependent. Thus, antiferromagnetism realized in layered form, such as in the A-type AFM order, is preferred to better couple with the field effect \cite{Burton:Prb,Dong:Prb13.2}. However, this type of AFM orders are rare in real materials. Although some manganites (e.g. LaMnO$_3$ and Nd$_{0.5}$Sr$_{0.5}$MnO$_3$) do display A-type AFM order \cite{Wollan:Pr,Dagotto:Prp}, the state is fragile and is not realized in thin films \cite{Bhattacharya:Prl,Dong:Prb08.3,Gibert:Nm}.

\begin{figure}
\centering
\includegraphics[width=0.46\textwidth]{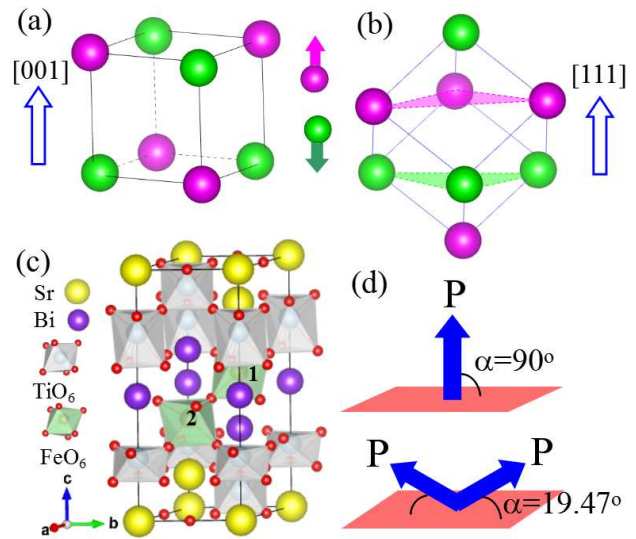}
\caption{(color online) (a-b) Sketches of G-type AFM order (as in BiFeO$_3$) viewed
from different orientations. The spins are distinguished by colors. (c) Sketch of a
superlattice stacking along the pseudo-cubic [$111$] direction. The two Fe's are labeled
as $1$ and $2$. (d) The possible orientations of $\textbf{P}$, with $\alpha$
being the angle between $\textbf{P}$ and the ($111$) plane.}
\label{F1}
\end{figure}

By contrast, the most common AFM state in pseudocubic perovskites is the G-type rocksalt-type order (shown in Fig.~\ref{F1}(a)). However, this G-type AFM order is actually layered along the pseudocubic [$111$] direction, as sketched in Fig.~\ref{F1}(b). From this observation, we propose the (BiFeO$_3$)$_{m}$/(SrTiO$_3$)$_{n}$ heterostructures grown along the [$111$] axis \cite{Xiao:Nc11} to realize the new $(\nabla\cdot\textbf{P})(\textbf{M}\cdot\textbf{L})$ ME coupling proposed here. There are several physical considerations to discuss:

First, BiFeO$_3$ is the most studied room-temperature multiferroic perovskite with prominent ferroelectricity
(a large $\textbf{P}$ up to $\sim90-100$ $\mu$C/cm$^{2}$ along the pseudo-cubic $[111]$ axes below a high $T_{\rm C}\sim1103$ K) \cite{Wang:Sci,Choi:Sci},
which is an advantage for realizations in field effects. The robust G-type AFM state of BiFeO$_3$ ($T_{\rm N}\sim643$ K) \cite{Wang:Sci} makes $\textbf{L}$
stable during the magnetoelectric switching.

Second, SrTiO$_3$ is the most used substrate, with various terminations and orientations available \cite{Gibert:Nm,Sanchez:Csr}. There is plenty of experience to fabricate BiFeO$_3$-SrTiO$_3$ heterostructures layer by layer along both the [001] and [111] orientations \cite{Wang:Sci,Heron:Apr,Yu:Ptrsa,Lu:Apr,Bruyer:Apl,Blok:Apl}. Moreover, the different valences between Sr$^{2+}$ and Bi$^{3+}$ can effectively modulate the interfacial carrier density, as in LaAlO$_3$-SrTiO$_3$ heterostructures \cite{Nakagawa:Nm}. Moreover, the electron transfer between BiFeO$_3$ and SrTiO$_3$ should be negligible due to the stability of the Fe$^{3+}$ and Ti$^{4+}$ ions, in contrast to the YFeO$_3$/YTiO$_3$ (or LaFeO$_3$/LaTiO$_3$) heterostructures where charge transfer occurs between Fe$^{3+}$ and Ti$^{3+}$ \cite{Kleibeuker:Prl,Zhang:Prb15}. In this sense, the BiFeO$_3$ layers are nearly perfectly isolated by SrTiO$_3$, as required.

Last but not least, because SrTiO$_3$ has a high dielectric constant \cite{Guo:Nl}, an applied voltage to the BiFeO$_3$-SrTiO$_3$ superlattice
will mainly affect the BiFeO$_3$ layers, making the electric switching of its $\textbf{P}$ possible.
In fact, a recent experiment has observed switchable ferroelectricity of BiFeO$_3$ bilayers sandwiched by SrTiO$_3$ layers \cite{Bruyer:Apl}.

\textit{Results \& Discussion}.- Standard density functional theory (DFT) calculations were performed to verify
the design proposed above~\cite{Supp}. First, a superlattice constructed from a BiFeO$_3$ bilayer and SrTiO$_3$
four-layer is studied, stacked along the pseudo-cubic [$111$] axis, as shown in Fig.~\ref{F1}(c). Here three layers
of Bi$^{3+}$, i.e. the double $n$-type interfaces, are adopted to dope one more electron to the Fe bilayer. The eight $[111]$
directions of $\textbf{P}$ can be classified into two groups: (i) two $\textbf{P}$'s pointing perpendicular to the interface
(up and down, or $\alpha=\pm90^{\circ}$); (ii) six $\textbf{P}$'s with an inclination relative to the interface
($\alpha=\pm19.47^{\circ}$), as summarized in Fig.~\ref{F1}(d). In the following, the $\alpha=\pm90^{\circ}$ cases
are the focus as the two end states of a FE switching process.

As summarized in Table~\ref{table1}, upon the FE switching, the local magnetic moments of the Fe ions show significant modulations, as a result of the carrier modulation of the field effect. Then, the net $\textbf{M}$ of the bilayer is switched from $-1$ $\mu_{\rm B}$ to $+1$ $\mu_{\rm B}$, accompanying the $\textbf{P}_{up}$ ($\alpha=+90^{\circ}$) to $\textbf{P}_{down}$ ($\alpha=-90^{\circ}$) switching \footnote{The ME coefficient, defined as $d\textbf{M}/d\textbf{E}$, can be roughly estimated as $|\Delta\textbf{M}|/E_c$ where the change of magnetization $|\Delta\textbf{M}|$ is about $155$ G (i.e. $0.5$ $\mu_{\rm B}$/Fe) and the coercive electric field $E_{c}$ for BiFeO$_3$ is of the order of $10$-$100$ kV/cm depending on material details \cite{Wang:Sci,Choi:Sci,Bruyer:Apl}. Thus $\alpha$ is estimated as $0.0155$-$0.00155$ Gcm/V, comparable to other ME heterostructures \cite{Duan:Prl2,Fechner:Prl}.}.

\begin{table}
\caption{DFT results. $\textbf{P}_{up}$ and $\textbf{P}_{down}$ denote the $\alpha=+90^{\circ}$ and $-90^{\circ}$ conditions, respectively. $m_{1}$ and $m_{2}$ are the local magnetic moments for the Fe$_{1}$ and Fe$_{2}$ cations, respectively, integrated within the Wigner-Seitz spheres. $M$ is the net bilayer magnetization. All moments in units of $\mu_{\rm B}$.}
\begin{tabular*}{0.48\textwidth}{@{\extracolsep{\fill}}lccc}
\hline \hline
FE & $m_1$ & $m_2$ & $M$\\
\hline
$\textbf{P}_{up}$ & $3.607$ & $-4.170$ & $-1$\\
$\textbf{P}_{down}$ & $4.170$ & $-3.608$ & $1$\\
\hline \hline
\end{tabular*}
\label{table1}
\end{table}

\begin{figure}
\centering
\includegraphics[width=0.48\textwidth]{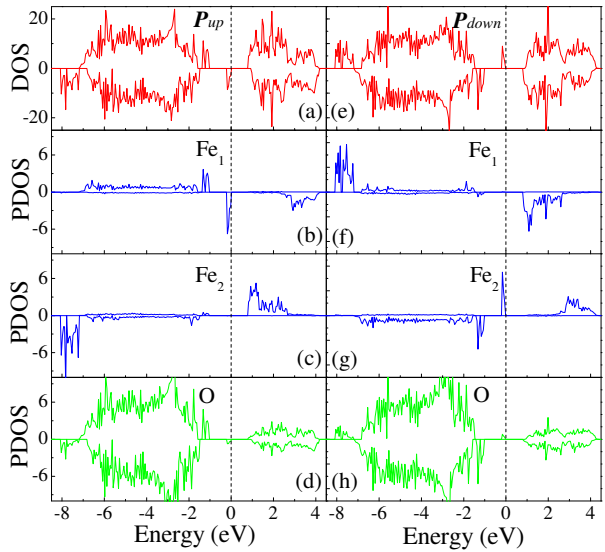}
\caption{(color online) Electronic structure (total DOS and pDOS) of the BiFeO$_{3}$/SrTiO$_{3}$
heterostructures along the [$111$] direction. Here, Bi trilayer and Fe bilayer are considered.
(a-d) $\textbf{P}_{up}$. (e-h) $\textbf{P}_{down}$. The Fermi energy is positioned at zero.}
\label{F2}
\end{figure}

In this heterostructure with a Bi-trilayer and a Fe-bilayer, one more electron is introduced into the system confined to the quantum well made by the Fe bilayer. Due to the field effect, the occupancy weight of the two Fe layers will be different. Moreover, the intrinsic tendency toward charge ordering will lead to the ideal
Fe$^{2+}$-Fe$^{3+}$ configuration, which gives rise to a $\pm1$ $\mu_{\rm B}$ net moment. Then, the FE switch will drive the switch between two magnetic-charge ordered configurations: Fe$^{3+}$(spin up)-Fe$^{2+}$(spin down) and Fe$^{2+}$(spin up)-Fe$^{3+}$(spin down) \cite{Supp}. This ideal limit indeed is confirmed by our DFT calculations, as revealed in the atomic-projected density of states (pDOS). As shown in Fig.~\ref{F2}, for Fe$_1$ the spin-down channel is occupied by one electron in the $\textbf{P}_{up}$ condition, i.e. Fe$^{2+}$, while the $3d$'s spin-down channel of Fe$_2$ is empty, i.e. Fe$^{3+}$. This ideal Fe$^{2+}$-Fe$^{3+}$ charge ordering also leads to insulating properties, compatible with the ferroelectricity of the BiFeO$_{3}$ bilayer.

\begin{figure}
\centering
\includegraphics[width=0.38\textwidth]{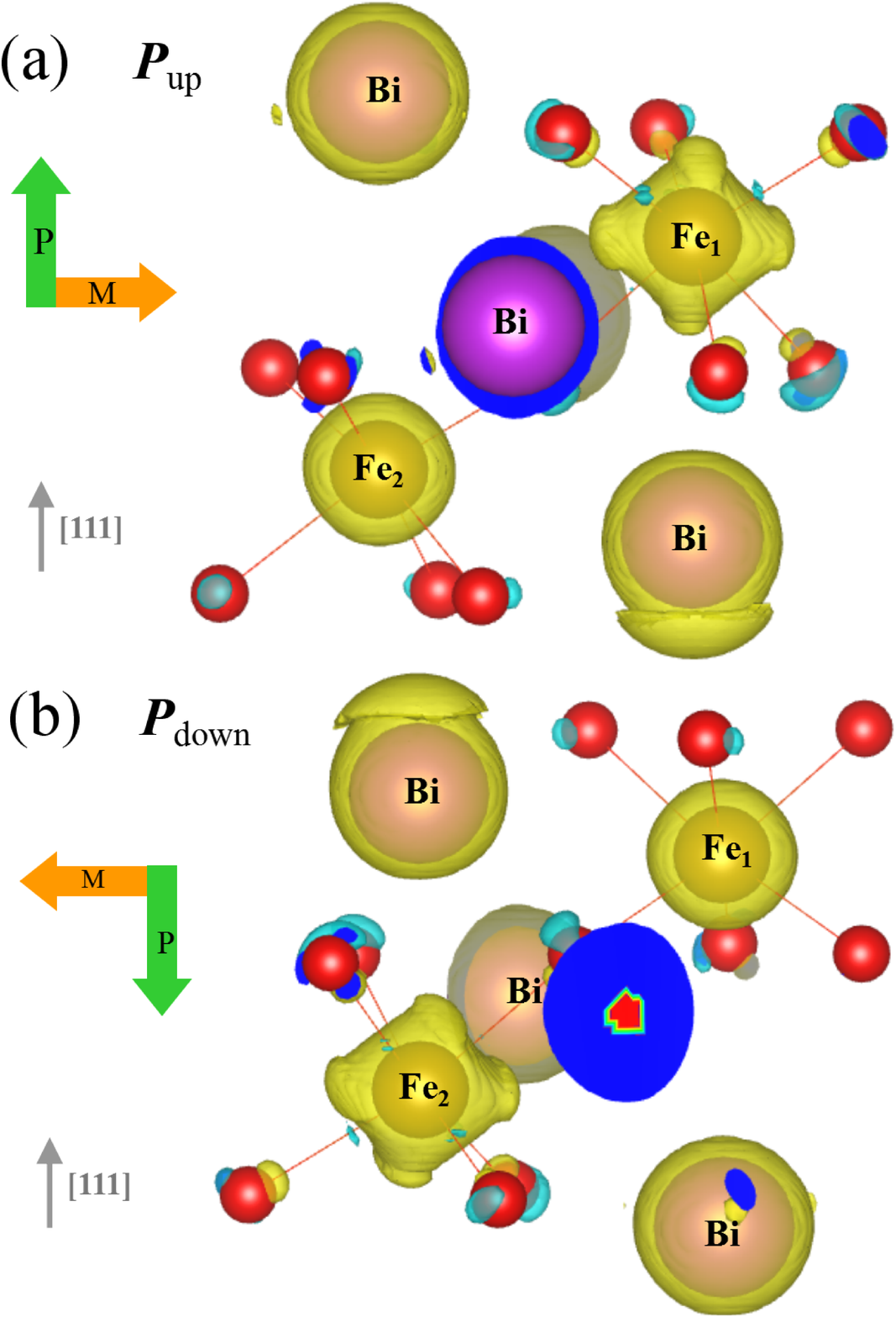}
\caption{(color online) Spatial distribution of the electronic density for the cases
(a) $\textbf{P}_{up}$ and (b) $\textbf{P}_{down}$. The orientations of $\textbf{M}$ and $\textbf{P}$ are also indicated.}
\label{F3}
\end{figure}

This FE switched charge ordering can be visualized by plotting the distribution of electrons (Fig.~\ref{F3}). First, the origin of ferroelectricity in the BiFeO$_3$ bilayer can be clearly seen as the bias of lone pair electrons of Bi$^{3+}$ ions. Second, the electron disproportion between Fe$_1$ and Fe$_2$ is very clear. The electron cloud surrounding the expected Fe$^{3+}$ ion is almost spherical, while for the Fe$^{2+}$ ion it is $d_{xz}$-shaped (or $d_{yz}$-shaped depending on the coordination choice) and larger in size.

\begin{figure}
\centering
\includegraphics[width=0.46\textwidth]{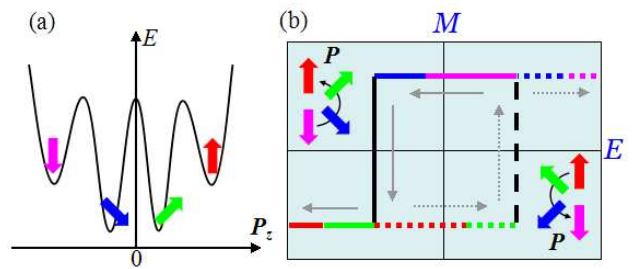}
\caption{(color online) (a) Sketch of energy {\it vs.} the $z$-axis component of $\textbf{P}$. (b) Sketch of the electric field control of magnetism. The sign of $\textbf{M}$ is turned accompanying the switch of $\textbf{P}$, forming an $\textbf{E}$-$\textbf{M}$ hysteresis loop. The maximum saturated $|\textbf{M}|$ can reach $0.5$ $\mu_{\rm B}$/Fe. The coercivity is determined by the FE coercivity of the BiFeO$_3$ layers. Even without the $\alpha=\pm90^\circ$ end states, a $\textbf{E}$-$\textbf{M}$ hysteresis loop can also be achieved between the $\alpha=\pm19.47^\circ$ cases. }
\label{F4}
\end{figure}

Besides the two end states, the intermediate states ($\alpha=\pm19.47^{\circ}$) are also calculated,
giving identical results to the corresponding $\alpha=\pm90^{\circ}$ limits (see Supplementary Materials).
In other words, the sign of the $c$-component of $\textbf{P}$ uniquely determines $\textbf{M}$, while the in-plane component does not affect this conclusion. This is reasonable considering the large spontaneous $\textbf{P}$ of BiFeO$_{3}$, whose $c$-component ($\sim30$ $\mu$C/cm$^{2}$) is already large enough for the field effect, even in the $\alpha=\pm19.47^{\circ}$ cases. The process leading to the complete electric-field switch of $\textbf{M}$ is summarized in Fig.~\ref{F4}, including an $\textbf{E}$-$\textbf{M}$ hysteresis loop, a desired function of magnetoelectricity.

Next, it is important to estimate the working temperature of this ME function. The approximate FE transition temperature $T_{\rm C}$ can be obtained by comparing the energy difference between the paraelectric and FE phases. As summarized in Table~\ref{table2}, the energy barrier is lowered by $29\%$ in bilayers compared with the bulk value. However, considering the very high FE $T_{\rm C}$ ($\sim1103$ K) of bulk BiFeO$_3$, the expected FE $T_{\rm C}$ of the BiFeO$_{3}$ bilayer should remain above room temperature, a favorable property.

To estimate the magnetic transition temperature $T_{\rm N}$, the exchange coefficient ($J$) is estimated by mapping the system to a classical spin model. In both bulk and bilayer systems, the nearest-neighbor $J$'s are AFM, leading to a G-type AFM state (Table~\ref{table2}). However, the magnitude of $J$ is reduced in bilayers, implying that the AFM coupling between Fe$^{2+}$-Fe$^{3+}$ is weaker than that between Fe$^{3+}$-Fe$^{3+}$. Considering the coordination number, the reduced dimensionality of bilayers will also suppress $T_{\rm N}$.

Another difference between bulk and bilayer is the magnetic anisotropy. For bulk BiFeO$_{3}$, with a spontaneous $\textbf{P}$ pointing along the hexagonal $z$-axis, the magnetic easy plane is the $x-y$ plane. In our DFT calculation with spin-orbit coupling (SOC), the magnetocrystalline energy is about $0.084$ meV/Fe, in agreement
with previous DFT results \cite{Ederer:Prb}. In fact, such a weak magnetic anisotropy is due to the Dzyaloshinskii-Moriya interaction, a high-order SOC effect, since the orbit moment of the high-spin $3d^{5}$ configuration is almost quenched. By contrast, in the BiFeO$_{3}$/SrTiO$_{3}$ heterostructures, a magnetocrystalline easy axis ($y$-axis) is found due to the spin-down $d_{xz}$ electron of Fe$^{2+}$, whose effective SOC is relatively large. Such a strong magnetocrystalline easy axis, rendering spins to be Ising-like, will be advantageous to increase $T_{\rm N}$. Using the coefficients (exchange and magnetocrystalline anisotropy, see Table~\ref{table2}) extracted from DFT calculations, a crude Monte Carlo (MC) simulation has been performed to estimate the phase transition temperatures \cite{Supp}. For the two end states ($\alpha=\pm90^{\circ}$), the simulated $T_{\rm N}$ of bilayers is about $\sim139$ K. However, this $T_{\rm N}$ can be significantly improved by using thicker Fe layers, as shown below.

\begin{table}
\caption{Summary of the calculated FE barrier $\Delta E$, exchange coefficient $J$ (with normalized spins $|\textbf{S}|=1$), and magnetocrystalline energy ($E_K$), all in units of meV/Fe. Three perpendicular spin axes (($x$, $y$, $z$), $z$: perpendicular to the bilayer) are adopted to calculate $E_K$,
and the energy for spins along the $z$-axis ($E_K(z)$) is taken as reference.}
\begin{tabular*}{0.48\textwidth}{@{\extracolsep{\fill}}ccccc}
\hline \hline
 & $\Delta E$  & $J$   & $E_K(x)$ & $E_K(y)$\\
\hline
bulk & $581$ & $39.72$& -0.084 & -0.084\\
\hline
$\alpha=\pm90^{\circ}$ & $414$ & $26.83$ & 0.165 & -0.250\\
\hline \hline
\end{tabular*}
\label{table2}
\end{table}

The calculations above have been done for ideal $3$ Bi plus $2$ Fe layers. In real superlattices, interfacial roughness may be present to some extend. It is necessary to check the stability of the above described ME function beyond the ideal conditions. To pursue this goal, both the layer numbers of Fe and Bi are changed to verify the ME function. Of course, the layer number of Fe must be even, or the net $\textbf{M}$ can not be flipped by the field effect. Then, besides the smooth interfaces, several hybrid cases with rough Bi layers have also been tested by using an (in-plane) doubling cell. As summarized in Table~\ref{table3}, it is clear that the only condition for the ME function is the nonstoichiometry between Bi and Fe, i.e. to have extra carriers no matter whether electrons or holes. In real experiments, even for those configurations with equivalent numbers of Bi and Fe layers (e.g. $2+2$), the proposed ME function remains valid once there is additional nonstoichiometry caused, e.g., by oxygen or Bi vacancies.

The ME function can also exist in thicker Fe layers, e.g. four Fe plus five Bi. Of course, the average $|\textbf{M}|$ per Fe will decrease with the thickness of Fe, since the inner Fe layers will not contribute to $\textbf{M}$ as much as the two interfacial layers. Even with this caveat, the thicker Fe cases can give rise to a moderate $|\textbf{M}|$ and more stable AFM order (unflipped during the ME switching), as well as enhanced $T_{\rm N}$, e.g. $\sim 371$ K for four Fe plus five Bi from the MC simulation \cite{Supp}, a favorable property.

\begin{table}
\caption{Validity of the proposed ME function in various conditions. All nonzero $\textbf{M}$ ($\mu_{\rm B}$/Fe units)
can be switched.}
\begin{tabular*}{0.48\textwidth}{@{\extracolsep{\fill}}cccccccc}
\hline \hline
Bi layer & $1$ & $2$ & $3$ & $1+2$ & $2+3$ & $1+3$ & $5$\\
\hline
Fe layer & $2$ & $2$ & $2$ & $2+2$ & $2+2$ & $2+2$ & $4$\\
\hline
$|\textbf{M}|$ & $0.5$ & $0$ & $0.5$ & $0.25$ & $0.25$ & $0$ & $0.23$ \\
\hline \hline
\end{tabular*}
\label{table3}
\end{table}

Let us reinterpret our DFT results in the context of the Landau theory. As stated before, the field effect can be represented by a nonzero $\nabla\cdot\textbf{P}$. Here, this field effect breaks the symmetry of the two end Fe layers. The bilayer AFM order parameter ($\textbf{L}$) can be expressed as $\textbf{M}_1-\textbf{M}_2$, where the subscript is the layer index. This order parameter $\textbf{L}$ is unchanged during the FE/magnetic switch, which is only determined by the initial condition. Considering the energy term $(\nabla\cdot\textbf{P})(\textbf{M}\cdot\textbf{L})$, the net magnetic moment $\textbf{M}$ can be switched accompanying the flipping of $\textbf{P}$, as proposed in the beginning. The only condition is that $|\textbf{M}|$ be nonzero, corresponding to a net ferrimagnetic moment ($\textbf{M}_1+\textbf{M}_2$) from the extra carriers (electrons or holes). Then, the phenomenological energy for the novel ME coupling can be described by:
\begin{eqnarray}
\nonumber F&\sim&(\nabla\cdot\textbf{P})(\textbf{M}\cdot\textbf{L})=(\nabla\cdot\textbf{P})[\textbf{M}_1^2-\textbf{M}_2^2].
\end{eqnarray}
Thus, our proposed ME function can be considered as a back coupling of two carrier-mediated ME interfaces.

Finally, note that some recent advances in ME heterostructures
reported the $180^\circ$ rotation of $\textbf{M}$ by electric fields
in metal/ferroelectric heterostructures \cite{Ghidini:Nc,Yang:Am}. However, the physical mechanism
relates with a process-dependent dynamics of magnetic moments
(a sequence of two $90^\circ$ $\textbf{M}$ rotations \cite{Hu:Nl}).
The primary driving force in these devices is the piezostrain modulated magnetocrystalline anisotropy, and usually
an assisting small magnetic field is needed \cite{Wang:Sr}. An alternative route is to tune the long-range
interaction (via the field effect) between two ferromagnetic layers separated by a nonmagnetic metal \cite{Fechner:Prl}.
Although pursuing a similar function, our design is conceptually different from these previous efforts.

\textit{Summary.}- To pursue the electric field control of magnetism, a new magnetoelectric coupling based on the field effect is here proposed, formally expressed as $(\nabla\cdot\textbf{P})(\textbf{M}\cdot\textbf{L})$. This new magnetoelectric coupling can realize the intrinsic $180^\circ$ flipping of magnetization accompanying the ferroelectric switching, while previously considered magnetoelectric couplings based on field effect can only modulate the magnetization amplitude. The new proposal is here predicted to be realized in practice using a few layers of BiFeO$_3$ ($111$) sandwiched in SrTiO$_3$. Benefiting from the robust G-type AFM state of BiFeO$_3$ and its prominent ferroelectricity, the net magnetization of BiFeO$_3$, of order $0.5$ $\mu_{\rm B}$/Fe, can be unambiguously switched by $180^\circ$ when flipping the ferroelectric polarization, leading to the expected $\textbf{E}$-$\textbf{M}$ hysteresis loop. Although only BiFeO$_3$ is studied here, our design principle based on $(\nabla\cdot\textbf{P})(\textbf{M}\cdot\textbf{L})$ can be extended to other magnetoelectric systems with polarization and antiferromagnetism, and may lead to practical magnetoelectric devices.

\begin{acknowledgments}
We acknowledge discussions with Hangwen Guo, Pu Yu, Xiaofang Zhai, Jinxing Zhang, and Junling Wang. This work was mainly supported by the National Natural Science Foundation of China (Grant Nos. 11274060 and 51322206), the Fundamental Research Funds for the Central Universities, Jiangsu Key Laboratory for Advanced Metallic Materials (Grant No. BM2007204), and Jiangsu Innovation Projects for Graduate Student (Grant No. KYLX15\underline{ }0112). E.D. was supported by the U.S. DOE, Office of Basic Energy Sciences, Materials Sciences and Engineering Division.
\end{acknowledgments}

\section{Supplementary material}

\section{Details of DFT method}
The density functional theory (DFT) calculations were performed based on the generalized gradient approximation (GGA) with revised Perdew-Burke-Ernzerhof (PBEsol) potentials \cite{Perdew:Prl08}, as implemented in the Vienna \textit{ab initio} Simulation Package (VASP) \cite{Kresse:Prb,Kresse:Prb96}. The cutoff energy of plane-wave is $550$ eV and the on-site Hubbard interaction ($U_{\rm eff}$) is imposed to Fe's $3d$ electrons using the Dudarev implementation \cite{Dudarev:Prb}. For BiFeO$_3$ bulk, a $7\times7\times3$ Monkhorst-Pack \textit{k}-point mesh centered at $\varGamma$ point is adopted for the Brillouin-zone integrations. For superlattices, the \textit{k}-point mesh is tuned correspondingly.

Cubic SrTiO$_3$ is chosen as the substrate with a lattice constant $3.900$ {\AA} (relaxed in DFT), which is very close to the experimental value $3.905$ {\AA}.

In our DFT calculations for superlattices, the initial absolute value of the magnetic moments of all Fe's are the same,
e.g. $4.5$ $\mu_{\rm B}$/$-4.5$ $\mu_{\rm B}$ or $5$ $\mu_{\rm B}$/$-5$ $\mu_{\rm B}$. The final results are robust i.e. they do not depend on the particular values used in these initial conditions.

\section{DFT results of BiFeO$_{3}$ bulk}
The ground state physical properties of BiFeO$_3$ bulk has been checked. Using GGA+$U$ calculation ($U_{\rm eff}=0, 2, 4$ eV tested), the structural parameters, including the cell volume $\Omega$, rhombohedral angle $\beta$, and atomic positions within the $R3c$ space group, are fully optimized and the calculated values are all in close agreement with the experimental values \cite{Kubel:Acb} and a previous study \cite{Neaton:Prb05}, as compared in Table~\ref{tableS1}. Particularly, for $U_{\rm eff}=4$ eV, the band gap is $2.0$ eV and the magnetic moment is $4.12$ $\mu_{\rm B}$ per Fe ion, which are also in agreement with experiments \cite{Gao:Apl06,Sosnowska:Apa}. The calculated polarization is about $88$ $\mu$C/cm$^{2}$ along the pseudocubic [111] axis, in reasonable agreement with the experimental value \cite{Li:Apl,Neaton:Prb05} as well. Thus, in this work $U_{\rm eff}=4$ eV on Fe ions will be adopted in the calculations for heterostructures.

\begin{table}
\footnotesize
\caption{Summary of DFT (GGA+$U$) results for BiFeO$_3$'s structural parameters in space group $R3c$. The Wyckoff positions are Bi($x$, $x$, $x$), Fe($x$, $x$, $x$), O($x$, $y$, $z$). The lattice constant of the rhombohedral primitive cell $a_{rh}$, the rhombohedral angle $\beta$, and unit cell volume $\Omega$ are provided. The data of ``Exp'' column is taken from experimental work \cite{Kubel:Acb}.}
\centering
\begin{tabular*}{0.46\textwidth}{@{\extracolsep{\fill}}cccccc}
\hline \hline
$U_{\rm eff}$ &  & 0 & 2 & 4 & Exp \\
\hline
  Bi & $x$ & 0.000 & 0.000 & 0.000 & 0.000 \\
  Fe & $x$ & 0.228 & 0.225 & 0.224 & 0.221 \\
  O & $x$ & 0.539 & 0.538 & 0.538 & 0.538 \\
   & $y$ & 0.941 & 0.940 & 0.940 & 0.933 \\
   & $z$ & 0.394 & 0.393 & 0.394 & 0.395 \\
\hline
  $a_{rh}$ ({\AA}) & \ & 5.56 & 5.60 & 5.60 & 5.63 \\
  $\beta$ ($^\circ$) & \ & 59.77 & 59.52 & 59.44 & 59.35 \\
  $\Omega$ ({\AA}$^{3}$) & \ & 121.04 & 122.54 & 122.86 & 124.32 \\
\hline \hline
\end{tabular*}
\label{tableS1}
\end{table}

\section{DFT results for superlattices}
\begin{figure}
\centering
\includegraphics[width=0.48\textwidth]{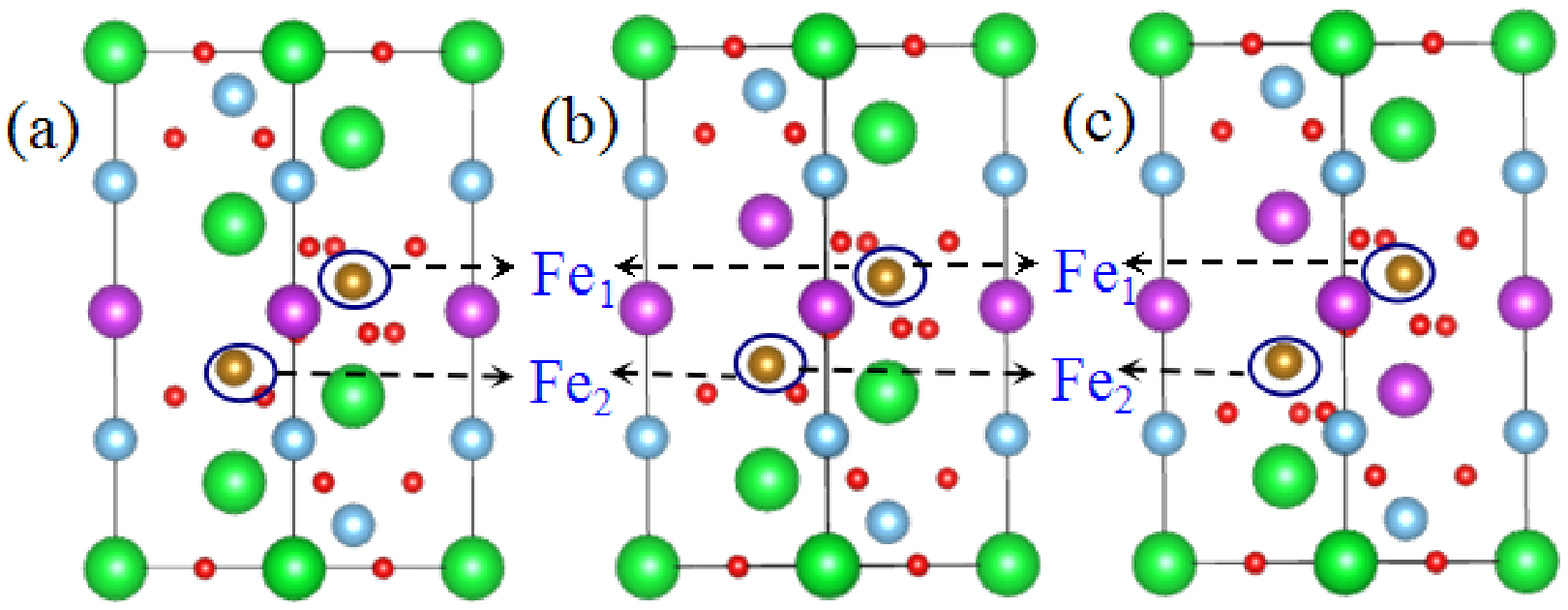}
\includegraphics[width=0.49\textwidth]{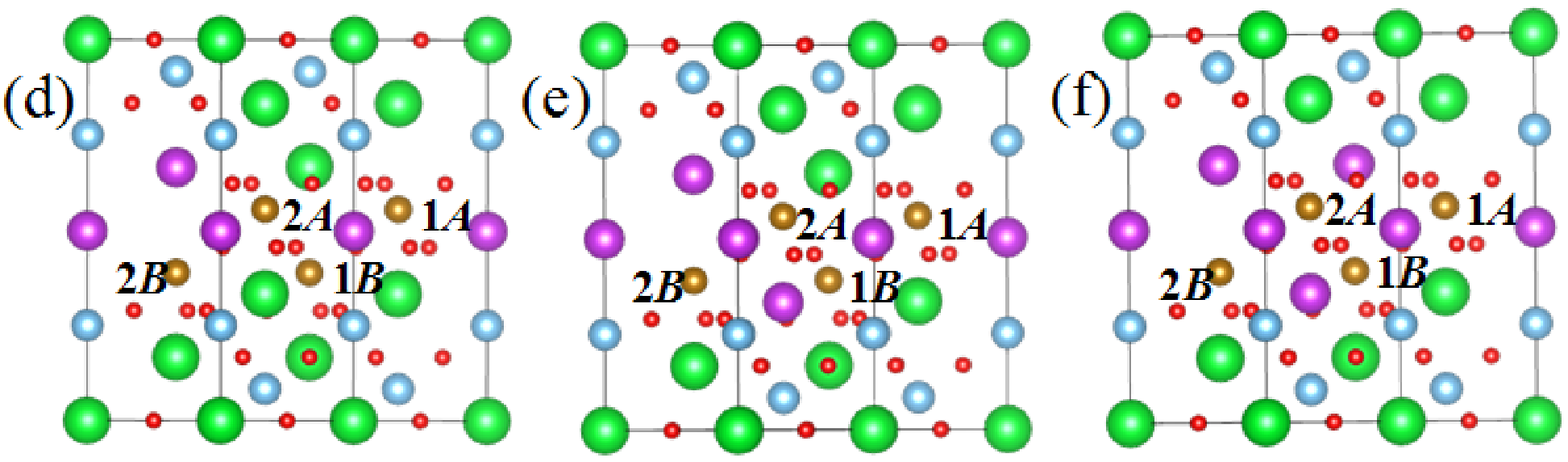}
\caption{(color online) BiFeO$_3$/SrTiO$_3$ superlattices. Here only Fe bilayers are displayed. (a) Bi monolayer; (b) Bi bilayer; (c) Bi trilayer.
(d-f) are for rough interfaces. (d) $1/2$ monolayer plus $1/2$ bilayer of Bi. (e) $1/2$ monolayer plus $1/2$ trilayer of Bi. (f) $1/2$ bilayer plus $1/2$ trilayer of Bi.}
\label{FS1}
\end{figure}

The crystalline structure of superlattices are shown in Fig.~\ref{FS1}, with different ratios of Bi and Fe. The possibility of rough interfaces is also considered.
The calculated magnetic moments are listed in Table~\ref{tableS2}. And the $\alpha=\pm19.47^{\circ}$ conditions always give rise to identical results (and thus not shown here) to the corresponding $\alpha=\pm90^{\circ}$ conditions. The local magnetic moments and net magnetization show significant modulations upon polarization switch except for stoichiometric cases. For asymmetric structures, exemplified in Fig.~\ref{FS1}(b), such a configuration will polarize the BiFeO$_3$ layers automatically because of the extra electric field induced by the asymmetric terminations. Thus, both the $\pm90^{\circ}$ initial polarization states will give an (almost) identical polarization direction after the atomic relaxation. In addition, in this case the net magnetization is zero because there are no extra carriers.
Therefore, this case is trivial regardless on whether the polarization can be reversed or not.

For symmetric terminations, independently of the thickness of Bi both the $\alpha=\pm90^{\circ}$ and $\alpha=\pm19.47^{\circ}$ polarized states are dynamically stable against structural relaxation, as occurs in BiFeO$_3$ bulk. Their energies are calculated and summarized in Table~\ref{tableS2}, and Fig. 4 in the main text is qualitatively sketched accordingly.

\begin{table}
\caption{The calculated energies $\Delta E$, local magnetic moments, and total magnetization for smooth interfaces. For each superlattice, the $\alpha=90^{\circ}$ state is considered as the energy of reference. $m_{1}$ and $m_{2}$ are the local magnetic moments for the Fe$1$ and Fe$2$ cations, respectively, integrated within the Wigner-Seitz spheres. $M$ is the net magnetization. All moments are in units of $\mu_{\rm B}$.}
\begin{tabular*}{0.48\textwidth}{@{\extracolsep{\fill}}cclcccr}
\hline \hline
Bi & Fe & $\alpha (^{\circ})$ & $\Delta E$ (meV) & $m_1$ & $m_2$ & $M$\\
\hline
$1$ & $2$ & $+90$ & $0$ & $4.147$ & $-3.522$ & $1$\\
  & & $+19.47$ & $-45$ & $4.163$ & $-3.486$ & $1$\\
  & & $-19.47$ & $-44$ & $3.486$ & $-4.163$ & $-1$\\
  & & $-90$ & $0$ & $3.522$ & $-4.147$ & $-1$\\
\hline
$3$ & $2$ & $+90$ & $0$ & $3.607$ & $-4.170$ & $-1$\\
  & & $+19.47$ & $-229$ & $3.628$ & $-4.178$ & $-1$\\
  & & $-19.47$ & $-229$ & $4.174$ & $-3.622$ & $1$\\
  & & $-90$ & $0$ & $4.170$ & $-3.608$ & $1$\\
\hline
$2$ & $2$ & $+90$ & $/$ & $4.151$ & $-4.162$ & $0$\\
  & &$-90$ & $/$ & $4.163$ & $-4.150$ & $0$\\
\hline
$5$ & $4$ & $+90$ & $0$ & & & $-0.95$\\
  & &$+19.47$ & $-900$ & & & $-0.99$\\
\hline \hline
\end{tabular*}
\label{tableS2}
\end{table}

\begin{table}
\caption{Local magnetic moments and total magnetization (in unit of $\mu_{\rm B}$) for rough interfaces. Here the thickness of Fe is fixed as bilayer. The subscripts A and B distinguish among the Fe ions in the same plane.}
\begin{tabular*}{0.48\textwidth}{@{\extracolsep{\fill}}cclccccr}
\hline \hline
& Bi & FE & $m_{1A}$ & $m_{1B}$ & $m_{2A}$ & $m_{2B}$ &$M$\\
\hline
$\alpha=\pm90^{\circ}$ & $1+2$ & $\textbf{P}_{up}$ & $4.148$ & $-4.164$ & $3.522$ & $-4.157$ & $-1$\\
& & $\textbf{P}_{down}$ & $4.157$ & $-3.512$ & $4.171$ & $-4.150$ & $1$\\
& $1+3$ & $\textbf{P}_{up}$ & $4.154$ & $-4.155$ & $4.154$ & $-4.166$ & $0$\\
& $2+3$ & $\textbf{P}_{up}$ & $4.145$ & $-4.155$ & $3.605$ & $-4.168$ & $-1$\\
& & $\textbf{P}_{down}$ & $4.160$ & $-4.155$ & $4.166$ & $-3.599$ & $1$\\
\hline \hline
\end{tabular*}
\label{tableS3}
\end{table}

\begin{figure}
\centering
\includegraphics[width=0.48\textwidth]{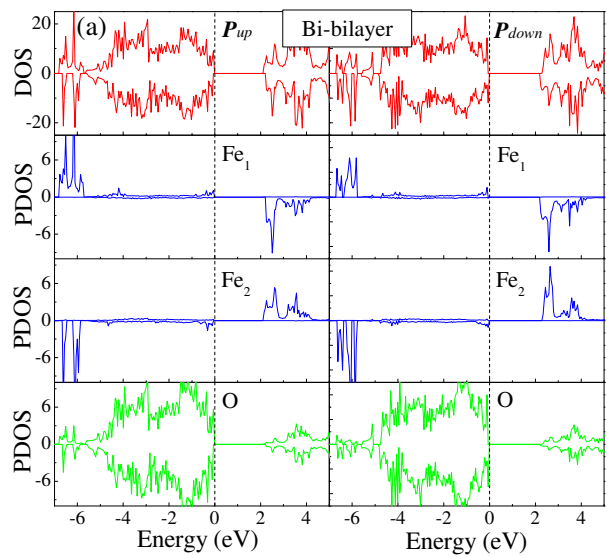}
\includegraphics[width=0.48\textwidth]{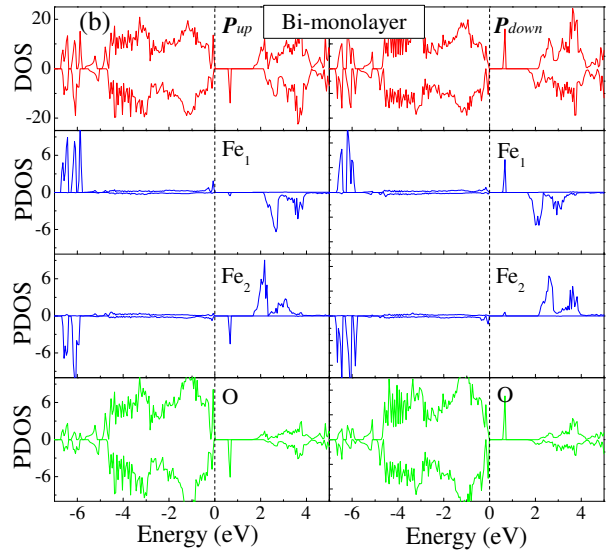}
\caption{(color online) Electronic structure (total DOS and pDOS) of BiFeO$_{3}$/SrTiO$_{3}$ superlattices. Only the $\alpha=\pm90^{\circ}$
conditions are presented (the $\alpha=\pm19.47^{\circ}$ conditions give very similar results). The Fermi energy is positioned at zero.
(a) Bi bilayer; (b) Bi monolayer. Both are for Fe bilayer, and the Bi trilayer is presented in the main text.}
\label{FS2}
\end{figure}

The atomic-projected density of states (pDOS) is presented in Fig.~\ref{FS2}. First, as shown in Fig.~\ref{FS2}(a), for Bi bilayer plus Fe bilayer both Fe1 and Fe2 are Fe$^{3+}$, with occupied $3d$ spin-up bands but empty spin-down bands. This configuration leads to zero net magnetization, which is trivial. Second, for Bi monolayer plus Fe bilayer, the extra carriers are holes. As illustrated in Fig.~\ref{FS2}(b), the in-gap unoccupied state is shared between Fe and oxygen, namely a hole is created from the Fe-O bonding state. This hole state leads to a nonzero moment $\textbf{M}=1$ $\mu_{\rm B}$.
This moment can also be switched by the polarization, rendering an identical magnetoelectric function as in the Bi trilayer case.

Similarly, for rough interfaces the local magnetic moments and net magnetization also show significant modulations upon
polarization switch except for the stoichiometric cases, as shown in Table~\ref{tableS3}.

\section{Exchange interaction \& magnetic anisotropy}
The Hamiltonian of the Heisenberg spin model with magnetocrystalline energy reads as:
\begin{equation}
H=\sum_{<\textbf{ij}>}J\textbf{S}_{\textbf{i}}\cdot\textbf{S}_{\textbf{j}}+\sum_{\textbf{i}}K_\textbf{i}(\textbf{S}_{\textbf{i}}\cdot\textbf{A}_\textbf{i})^2,
\end{equation}
where $J$ is the exchange interaction between the nearest-neighbor spins $\textbf{S}_{\textbf{i}}$ and $\textbf{S}_{\textbf{j}}$; $K_{\textbf{i}}$ is the coefficient of magnetic anisotropy and $\textbf{A}$ is a unit vector along the magnetocrystalline axis. According to the energy differences between antiferromagnetism and ferromagnetism, the exchange coefficients $J$ can be estimated with normalized spins ($|\textbf{S}|=1$). By fixing the antiferromagnetic configuration and enabling the spin-orbit coupling, the magnetocrystalline coefficients and axes can be calculated by rotating the spins' directions. The results are summarized in Table~\ref{tableS4}. For BiFeO$_3$ bulk, the magnetic easy plane is the $x$-$y$ plane, where the $z$-axis is the polarization direction. This magnetocrystalline anisotropy is mainly due to the Dzyaloshinskii-Moriya interaction, a higher order effect of the spin-orbit coupling. In more detail, the Fe$^{3+}$ ion should have a weak magnetocrystalline anisotropy [$K^{3+}$, $\textbf{A}^{3+}$=($0$, $0$, $1$)].
By contrast, in the superlattice ($\alpha=\pm90^{\circ}$), a magnetocrystalline easy axis ($y$-axis) is found due to the spin-down $d_{xz}$ electron of Fe$^{2+}$. Thus, Fe$^{2+}$ has two possible sources of magnetocrystalline anisotropy i.e. [$K^{3+}$, $\textbf{A}^{3+}$=($0$, $0$, $1$)] plus [$K^{xz}$, $\textbf{A}^{xz}$=($0$, $1$, $0$)].

\begin{table}
\caption{Summary of the exchange $J$ and magnetocrystalline coefficients $K$ found both for BiFeO$_{3}$ bulk and the superlattice (Bi-trilayer plus Fe-bilayer) in the $\alpha=\pm90^{\circ}$ condition. The unit is meV/Fe.}
\begin{tabular*}{0.48\textwidth}{@{\extracolsep{\fill}}cccc}
\hline \hline
 & $J$ & $K$ & \textbf{A} \\
\hline
bulk & $39.72$ & $0.084$ & ($0$, $0$, $1$)\\
\hline
Fe$^{3+}$ in bilayer &  & $-0.165$ & ($0$, $0$, $1$)\\
Fe$^{2+}$ in bilayer & & $-0.846$ & ($0$, $0.981$, $0.195$)\\
Fe$^{2+}$-Fe$^{3+}$ & $26.83$ & &\\
\hline \hline
\end{tabular*}
\label{tableS4}
\end{table}

\begin{figure}
\centering
\includegraphics[width=0.4\textwidth]{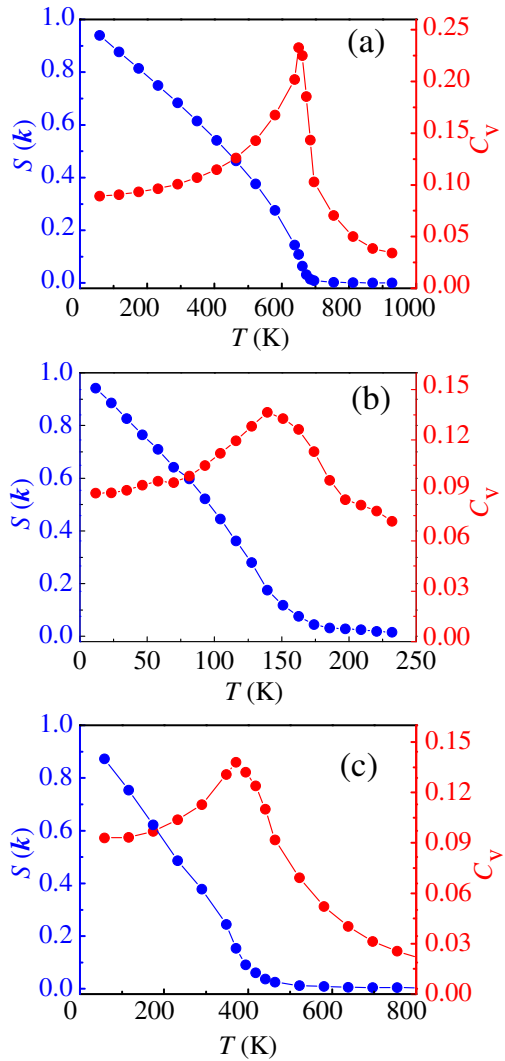}
\caption{(color online) Monte Carlo results for the spin structure factor $S({\textbf{k}})$ and specific heat $C_{\rm v}$ as a function of temperature. (a) BiFeO$_3$ bulk (size $L\times L\times L$, $L=20$); $\textbf{k}$=($\pi$, $\pi$, $\pi$). (b) Fe bilayer (size $L\times L$, $L=20$), $\textbf{k}$=($\pi$, $\pi$). The magnetic coefficients listed in Table~\ref{tableS4} are used. (c) Four Fe plus five Bi.}
\label{FS3}
\end{figure}

\section{Monte Carlo Simulations}
To obtain the magnetic transition temperature ($T_{\rm N}$) of the Fe bilayer, the Heisenberg spin model with periodic boundary conditions is studied and the standard Markov chain Monte Carlo (MC) method with the Metropolis algorithm is employed to investigate phase transitions. In our MC simulation, the first $4\times10^4$ MC steps are employed for thermal equilibrium while the following $1\times10^4$ MC steps are used for measurements. In all simulations, the acceptance ratio of MC updates is controlled to be about $50\%$ by adjusting the updating windows for spin vectors.
The specific heat per site ($C_{\rm v}$($T$)) is measured as a function of temperature ($T$). $C_{\rm v}$($T$) is calculated using the standard fluctuation equation: $N(<E^2>-<E>^2)/k_{\rm B}T^2$, where $k_{\rm B}$ is the Boltzmann constant, $N$ is the number of total sites, and $<>$ denotes the MC average.

To characterize the different magnetic orders, the spin structure factor is also calculated, which reads as \cite{Dong:Prb08,Dong:Prb08.2}:
\begin{equation}
S(\textbf{k})=\frac{1}{N^2}\sum_{\textbf{j}}\textbf{S}_{\textbf{j}}\cdot\textbf{S}_{\textbf{j}+\textbf{r}}\exp[i\textbf{k}\cdot(\textbf{j}+\textbf{r})],
\end{equation}
where $\textbf{r}$ and $\textbf{k}$ are vectors in real and reciprocal spaces, respectively.

As shown in Fig.~\ref{FS3}, both the spin structure factors and specific heats present an antiferromagnetic phase transition at $656\pm6$ K for the bulk (very close to the experimental value $\sim643$ K), and
a ferrimagnetic transition at $145\pm6$ K for the Bi trilayer plus Fe bilayer
with the $\alpha=\pm90^\circ$ conditions, as well as a higher ferrimagnetic transition at $383\pm12$ K
for five Bi plus four Fe with the $\alpha=\pm90^\circ$ conditions.

\section{Feasibility of superlattices}

\begin{figure}
\centering
\includegraphics[width=0.5\textwidth]{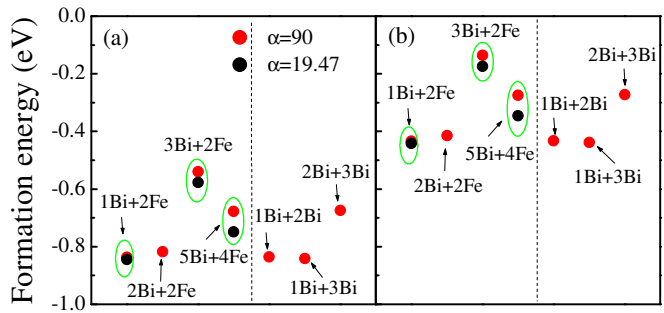}
\caption{The formation energies (per $AB$O$_3$ u.c.) of BiFeO$_3$/SrTiO$_3$ superlattices calculated by comparing their energies with equal amounts of: (a) Bi$_2$O$_3$, Fe$_2$O$_3$, SrO, O$_2$, and TiO$_2$; or (b) BiFeO$_3$, SrTiO$_3$, Bi$_2$O$_3$, SrO, and O$_2$. In all cases the formation energies of our designs are highly negative, implying the energetic stability of superlattices.}
\label{FS4}
\end{figure}

To study the feasibility of the proposed superlattices, the corresponding formation energies are estimated by comparing
the energies of the superlattices with equal amounts of (a) Bi$_2$O$_3$, SrO, O$_2$, Fe$_2$O$_3$, TiO$_2$,
or (b) Bi$_2$O$_3$, SrO, O$_2$, BiFeO$_3$, SrTiO$_3$.
In both cases (a) and (b), the relative formation energies of our designs are all highly negative,
as shown in Fig.~\ref{FS4}, implying their potential energetic stability.

In practice, several factors can affect the growth of films, such as temperature, type of substrate, atmosphere conditions, sources, methods, and many others. Moreover, usually these oxide heterostructures are fabricated (via PLD, MBE, or other methods) at high temperatures (and thus high energies) with a passive and kinetic growth processes
(e.g. two dimensional layer-by-layer deposition). As a consequence, the ground state energies are not really
the key factor to consider to analyze superlattice stabilities. Many superlattices are not in the lowest energy configurations, but they can still be fabricated and they are stable. For example, the (LaMnO$_3$)$_n$/(LaNiO$_3$)$_n$ superlattices are available for various integers $n$'s and also for different stacking orientations \cite{Gibert:Nm}, although there must be only one that has the lowest energy. Many other experimental examples are also available such as LaMnO$_3$/SrMnO$_3$, LaFeO$_3$/LaCrO$_3$, BiFeO$_3$/La$_{0.7}$Sr$_{0.3}$MnO$_3$, LaAlO$_3$/SrTiO$_3$, etc. The energy, entropy, chemical potentials, interface, growth sequence, and many other factors will co-determine the success of a sample growth. Therefore, in principle it is impossible to reject {\it a priori} a proposed superlattice without trying its growth merely based on energetic considerations.

A recent experiment has demonstrated that a BiFeO$_3$ film can be deposited on a SrTiO$_3$ ($111$) substrate in the two-dimensional growth mode once a buffer layer SrRuO$_3$ is added \cite{Blok:Apl}, providing a key technical advance to prepare our designs. In addition, the technique of termination control has also been realized for perovskite oxide substrates and BiFeO$_3$/La$_{0.7}$Sr$_{0.3}$MnO$_3$ heterostructures on a SrTiO$_3$ ($001$) substrate \cite{Yu:Pnas,Sanchez:Csr}. Therefore, there are neither fundamental problems nor physical concerns that prevent the fabrication of our designed superlattices with modern instruments and state-of-art techniques.
Our present work will further stimulate experimentalists to study the proposed BiFeO$_3$ ($111$) films.

\section{Switchable ferroelectricity}

Early experiments showed that the polarization could be switched in ultrathin BiFeO$_3$ films even down to $4$-$5$ u.c. \cite{Chu:Apl,Bea:Jjap,Maksymovych:Prb}.
A recent experiment found switchable ferroelectricity in BiFeO$_3$ bilayers sandwiched in SrTiO$_3$ layers \cite{Bruyer:Apl}. These results were obtained for the [$001$]-oriented  BiFeO$_3$ with a surface or asymmetric terminations.

An asymmetric interface/surface will self-pole the BiFeO$_3$ layers and bias the ferroelectric hysteresis loop. In the worst condition, this self-poling effect could be too strong, making the polarization not switchable, like the case shown in Fig.~\ref{FS1}(b). However, for superlattices with symmetric interfaces, like the cases shown in Fig.~\ref{FS1}(a) and Fig.~\ref{FS1}(c), this self-poling effect can be reduced to a minimal level, if not ideally zero, and thus the ferroelectric polarization can be switched even down to one Fe layer (and two Bi layers). In fact, a milestone experimental study has proved before that there are no thickness limits imposed on practical devices by an intrinsic ferroelectric size effect \cite{Fong:Sci}.

The break-down field may be reduced when the BiFeO$_3$ layer is ultra-thin. Meanwhile, the ferroelectric energy barrier is also lowered, as revealed in our calculation, which means the required switching field is also lowered. A recent experiment on a (BiFeO$_3$)$_2$/(SrTO$_3$)$_4$ superlattice has confirmed the switching ability, with indeed a lowered coercive field (which is an advantage for the magnetoelectric coefficient) \cite{Bruyer:Apl}.

\section{Switch of $\textbf{M}$ {\it vs} $\textbf{L}$}

If $\textbf{L}$ is flipped during the $\textbf{P}$ switching, then $\textbf{M}$ will not be switched. Having an unchanged $\textbf{L}$ is indeed an assumption of our current study. It is indeed a technical challenging open question in the theoretical analysis of this type of problems to rigorously simulate this {\it dynamical} magnetoelectric process. However, it is reasonable to assume that $\textbf{L}$ will remain unchanged since it is very difficult to flip synchronously all spins by $180^{\circ}$ in a G-type antiferromagnet. Even if the synchronous flip of a pair Fe(up)-Fe(down) to Fe(down)-Fe(up) could occur locally in the bilayer case, it is not a serious problem.

First, the $5$ Bi plus $4$ Fe case (and other thicker cases with even Fe layers but odd Bi layers)
can show the same magnetoelectric switching function although the magnetic moment per Fe is smaller
(but still considerable large). It is difficult to imagine that the ferroelectric switching
can flip four (or more) layer spins (up-down-up-down-...) by $180^{\circ}$ synchronously.
In fact, we recommend to study $5$ Bi plus $4$ Fe in future experiments to pursue a room temperature function.

Second, for real functional devices based on our designs, the polarization will not be switched globally, but in a very small area (i.e. one-bit by one-bit operation). Clearly the global antiferromagnetic order cannot be flipped by such a local ferroelectric switching. In addition, the local flip of $\textbf{L}$ is also unlikely, because it will generate antiferromagnetic domain walls that will cost considerable energy because of the large exchange coupling between Fe magnetic moments.

Finally, even if $\textbf{L}$ (instead of $\textbf{M}$) could be flipped by ferroelectric switching in some cases, it would open another interesting topic related to electromagnons (certainly beyond the current work and deserving of independent studies).

In summary for the issue of antiferromagnetism, it is reasonable to assume that $\textbf{L}$ will remain unchanged, especially for the thicker BiFeO$_3$ cases. As long as $\textbf{L}$ is unchanged, the physical quantity $\textbf{M}$ must be flipped accompanying the $\textbf{P}$ switching according to the symmetry requirements as well as microscopic driven force.

\section{Measurement of $\textbf{M}$}
With regards to the experimental measurement of the magnetic configuration discussed here, macroscopically this magnetic configuration can be verified via standard magnetic measurements. For example, for the bilayer case the saturated moment should be $0.5$ $\mu_{\rm B}$/Fe. For thicker cases, the moment per Fe will decrease, but still it will be sufficiently large to be measured. For example, a commercial SQUID can detect such a moment if a superlattice is fabricated
with a thickness of a dozen nm. Microscopically, the proposed magnetic configuration can be detected using neutron techniques, such as PNR (polarized neutron reflectivity), or via XMCD (X-ray magnetic circular dichroism).

\bibliographystyle{apsrev4-1}
\bibliography{ref}
\end{document}